\documentclass[12pt,a4paper,twoside]{article}
\usepackage{graphics}
\usepackage{delarray}
\newlength{\figwidth}
\newcommand{\eqref}[1]{(\ref{#1})}
\title{ 
\vspace{-2cm}
  {\large\sc   Institute of Experimental Physics} \\[-6mm]
  {\large \sc  Warsaw University} 
\vspace{-0.5cm}
\begin{flushright}
 {\large hep-ph/0006335 } 
\end{flushright}
\vspace{0.5cm}
Leptoquark Searches at Future Colliders }
\author{Aleksander Filip \.Zarnecki \\
{\small\it 
Institute of Experimental Physics, 
Warsaw University,
Ho\.za 69, 00-681 Warszawa, Poland} \\
{\small\it E-mail: zarnecki@fuw.edu.pl}
} 
\begin{document} 

\maketitle 

\begin{abstract}
Leptoquark searches are considered for future experiments at HERA, 
Tevatron, LHC, TESLA and THERA.
Expected exclusion limits from direct leptoquark searches
(leptoquark pair production and single leptoquark production)
are compared with indirect limits expected from the contact interaction
analysis.
Strongest limits on the leptoquark masses and couplings are
expected from high luminosity LHC data.
If the leptoquark signal suggested by the existing data is detected at LHC, 
with mass below 1 TeV, TESLA will be an ideal place to study leptoquark
properties, provided that the leptoquark Yukawa coupling is not too small.
\end{abstract}

\thispagestyle{empty}

%
%

\section{Introduction}
\label{sec-intro}

New result on atomic parity violation (APV) in Cesium and 
unitarity of the CKM matrix, as well as recent LEP2 hadronic 
cross-section measurements indicate possible 
deviations from the Standard Model predictions. 
Exchange of leptoquark type objects with masses above 250GeV 
has been proposed as a possible explanation for these effects \cite{myglqa}. 
If the observed signal is real it should
become clearly visible in future colliders.

The aim of the present analysis is to compare leptoquark
search limits expected from different experiments.
The Buchm\"uller-R\"uckl-Wyler model used in this
analysis is described in  section \ref{sec-model}. 
Results from the global analysis of available experimental data\cite{myglqa}
and the possible leptoquark signal are briefly summarized in
section \ref{sec-cur}.
Parameters of existing and future colliders considered in this analysis are
presented in section \ref{sec-exp}.

In section \ref{sec-ci} expected limits on the leptoquark mass 
to the coupling ratio are calculated in the limit of very high 
leptoquark masses, using the contact interaction approximation.
Limits expected from precise measurements of the Drell-Yan lepton 
pair production at the Tevatron and LHC, hadronic cross-section 
measurements at TESLA $e^{+}e^{-}$ and high-$Q^{2}$ NC DIS at HERA, 
THERA and TESLA ($e^{+}e^{-}$ and $e\gamma$) are compared with
the existing limits.
In sections \ref{sec-pair} and \ref{sec-single} expected limits from
direct leptoquark production are considered.
Leptoquark pair production is studied  for the Tevatron, LHC and TESLA 
($e^{+}e^{-}$ and $\gamma\gamma$ scattering)
whereas  single leptoquark production is considered for HERA, THERA and
TESLA ($e^{+}e^{-}$ and $e\gamma$ scattering).
Results expected from different experiments are compared
in section \ref{sec-comp}.


\section{Leptoquark models}
\label{sec-model}

In this paper a general classification of leptoquark states 
proposed by Buchm\"uller, R\"uckl and Wyler \cite{brw} will be used.
The Buchm\"uller-R\"uckl-Wyler (BRW) model is based on 
the assumption that new interactions should respect the 
$SU(3)_{C} \times SU(2)_{L} \times U(1)_{Y}$ symmetry of
the Standard Model.
In addition leptoquark couplings are assumed to be family diagonal
(to avoid FCNC processes) and to conserve lepton and baryon numbers
(to avoid rapid proton decay).
Taking into account very strong bounds from rare decays
it is also assumed that leptoquarks couple either to left-
or to right-handed leptons.
With all these assumptions there are 14 possible states 
(isospin singlets or multiplets) of scalar and vector leptoquarks.
Table \ref{tab-aachen} lists these states according to 
the so-called Aachen notation \cite{aachen}.
An S(V) denotes a scalar(vector) leptoquark and the subscript
denotes the weak isospin.
When the leptoquark can couple to both right- and left-handed
leptons, an additional superscript indicates  the lepton chirality.
A tilde is introduced to differentiate between leptoquarks
with different hypercharge.
\begin{table}[tbp]
  \begin{center}
   \begin{tabular}{lcccccc}
      \hline\hline\hline\noalign{\smallskip}
Model & Fermion & Charge & $BR(LQ \rightarrow e^{\pm}q)$ & 
        \multicolumn{2}{c}{Coupling} & Squark \\
      & number F &   Q   & $\beta$  &   &  & type \\
\hline\hline\hline\noalign{\smallskip}
$S_{\circ}^L$ &  2  &  $-1/3$  &  1/2  &   $e_{L}u$ & $\nu d$  & $\tilde{d_R}$ \\
\hline\noalign{\smallskip}
$S_{\circ}^R$ &  2  &  $-1/3$  &  1  &  $e_{R}u$ &  &   \\
\hline\noalign{\smallskip}
$\tilde{S}_{\circ}$  &  2  &  $-4/3$   &  1  &  $e_{R}d$ & &    \\
\hline\noalign{\smallskip}
$S_{1/2}^L$   &  0  &  $-5/3$   &  1  &  $e_{L} \bar{u}$ & &  \\
              &     &  $-2/3$   &  0  &  & $\nu \bar{u}$  &   \\
\hline\noalign{\smallskip}
$S_{1/2}^R$   &  0  &  $-5/3$  &  1 &  $e_{R} \bar{u}$  &  &  \\
              &     &  $-2/3$  &  1 &  $e_{R} \bar{d}$  &  &  \\
\hline\noalign{\smallskip}
$\tilde{S}_{1/2}$ &  0 &  $-2/3$  &  1  &  $e_{L} \bar{d}$  &  & $\overline{\tilde{u}_{L}}$  \\
                  &    &  $+1/3$  &  0  &     & $\nu \bar{d}$  & $\overline{\tilde{d}_{L}}$  \\
\hline\noalign{\smallskip}
$S_{1}$       &  2  & $-4/3$  &  1  & $e_{L}d$ & & \\
              &     & $-1/3$  &  1/2 & $e_{L}u$ &  $\nu d$  & \\
              &     & $+2/3$  &  0   &  &  $\nu d$  &  \\
\hline\hline\hline\noalign{\smallskip}
$V_{\circ}^L$ &  0  &  $-2/3$  &  1/2  &   $e_{L}\bar{d}$ & $\nu \bar{u}$ & \\
\hline\noalign{\smallskip}
$V_{\circ}^R$ &  0  &  $-2/3$  &  1  &  $e_{R}\bar{d}$ & & \\
\hline\noalign{\smallskip}
$\tilde{V}_{\circ}$  &  0  &  $-5/3$   &  1  &  $e_{R}\bar{u}$  & &  \\
\hline\noalign{\smallskip}
$V_{1/2}^L$   &  2  &  $-4/3$   &  1  &  $e_{L} d$  & & \\
              &     &  $-1/3$   &  0  &  & $\nu d$  &   \\
\hline\noalign{\smallskip}
$V_{1/2}^R$   &  2  &  $-4/3$  &  1 &  $e_{R} d$  &  &   \\
              &     &  $-1/3$  &  1 &  $e_{R} u$  &  &  \\
\hline\noalign{\smallskip}
$\tilde{V}_{1/2}$ &  2 &  $-1/3$  &  1  &  $e_{L} u$ &  &  \\
                 &     &  $+2/3$   &  0  &  & $\nu u$  &   \\
\hline\noalign{\smallskip}
$V_{1}$       &  0  & $-5/3$  &  1  & $e_{L} \bar{u}$ & \\
              &     & $-2/3$  &  1/2 & $e_{L}\bar{d}$ & $\nu \bar{u}$ & \\
              &     & $+1/3$  &  0   &   &  $\nu \bar{d}$  &  \\
\hline\hline\hline\noalign{\smallskip}
    \end{tabular}
  \end{center}
  \caption{A general classification of leptoquark states 
 in the Buchm\"uller-R\"uckl-Wyler model. 
 Listed are the leptoquark fermion number, F, 
 electric charge, Q (in units of elementary charge), 
the branching ratio to electron-quark (or electron-antiquark), $\beta$
 and the flavours of the coupled lepton-quark pairs. 
Also shown are possible squark assignments to the leptoquark states
 in  the minimal supersymmetric theories with broken R-parity.}
  \label{tab-aachen}
\end{table}
Listed in Table \ref{tab-aachen} are the leptoquark fermion
number F, electric charge Q, and the branching ratio to an electron-quark
pair (or electron-antiquark pair), $\beta$.
The leptoquark branching fractions are predicted by the BRW model 
and are either 1, $\frac{1}{2}$ or 0.
For a given electron-quark branching ratio $\beta$, the branching ratio 
to the neutrino-quark is by definition $(1-\beta)$. 
Also included in Table \ref{tab-aachen} are the flavours and chiralities of 
the lepton-quark pairs coupling to a given leptoquark type.
In three cases the squark flavours (in supersymmetric theories with 
broken R-parity) with corresponding  couplings
are also indicated.
Present analysis takes into account only leptoquarks which couple
to the first-generation leptons ($e$, $\nu_{e}$) and first-generation 
quarks ($u$, $d$), as most of the existing experimental data  
constrain this type of couplings.
Second- and third-generation leptoquarks as well as generation-mixing
leptoquarks will not be considered in this paper.
It is also assumed that one of the leptoquark 
types gives the dominant contribution, as compared with other leptoquark
states 
and that the interference between different leptoquark states can be neglected.
Using this simplifying assumption, different leptoquark types 
can be considered separately.
Finally, it is assumed that different leptoquark states within 
isospin doublets and triplets have the same mass.


\section{Current limits from global analysis}
\label{sec-cur}

In a recent paper\cite{myglqa} available data from HERA, LEP 
and the Tevatron, as well as from low energy  
experiments are used to constrain the Yukawa couplings $\lambda_{LQ}$
and masses $M_{LQ}$ for scalar and vector leptoquarks.
To compare the data with predictions of the BRW model
the global probability  function ${\cal P}(\lambda_{LQ},M_{LQ})$ 
is introduced, describing the probability that the data come from
the model described by parameters $\lambda_{LQ}$ and $M_{LQ}$. 
The probability  function  is defined in such a way that the Standard Model 
probability ${\cal P}_{SM} \equiv 1$.
Constraints on the leptoquark couplings and masses were
studied in the limit of very high leptoquark masses
(using the contact interaction approximation \cite{lqci})
as well as for finite leptoquark masses, with mass effects 
correctly taken into account.
Excluded on  95\% confidence level are all models (parameter values) 
which result in  the global probability
less than 5\% of the Standard Model probability:
${\cal P}(\lambda_{LQ},M_{LQ}) < 0.05$.
For models which describe the data
much better than the Standard Model
(${\displaystyle {\cal P}_{max} \; \equiv \;
\max_{\lambda,M} {\cal P}(\lambda,M) \gg 1}$)
the 95\% CL  signal limit is defined by the condition:
${\cal P}(\lambda_{LQ},M_{LQ}) > 0.05 \cdot {\cal P}_{max}$.

\begin{figure}[tbp]
\centerline{\resizebox{\figwidth}{!}{%
  \includegraphics{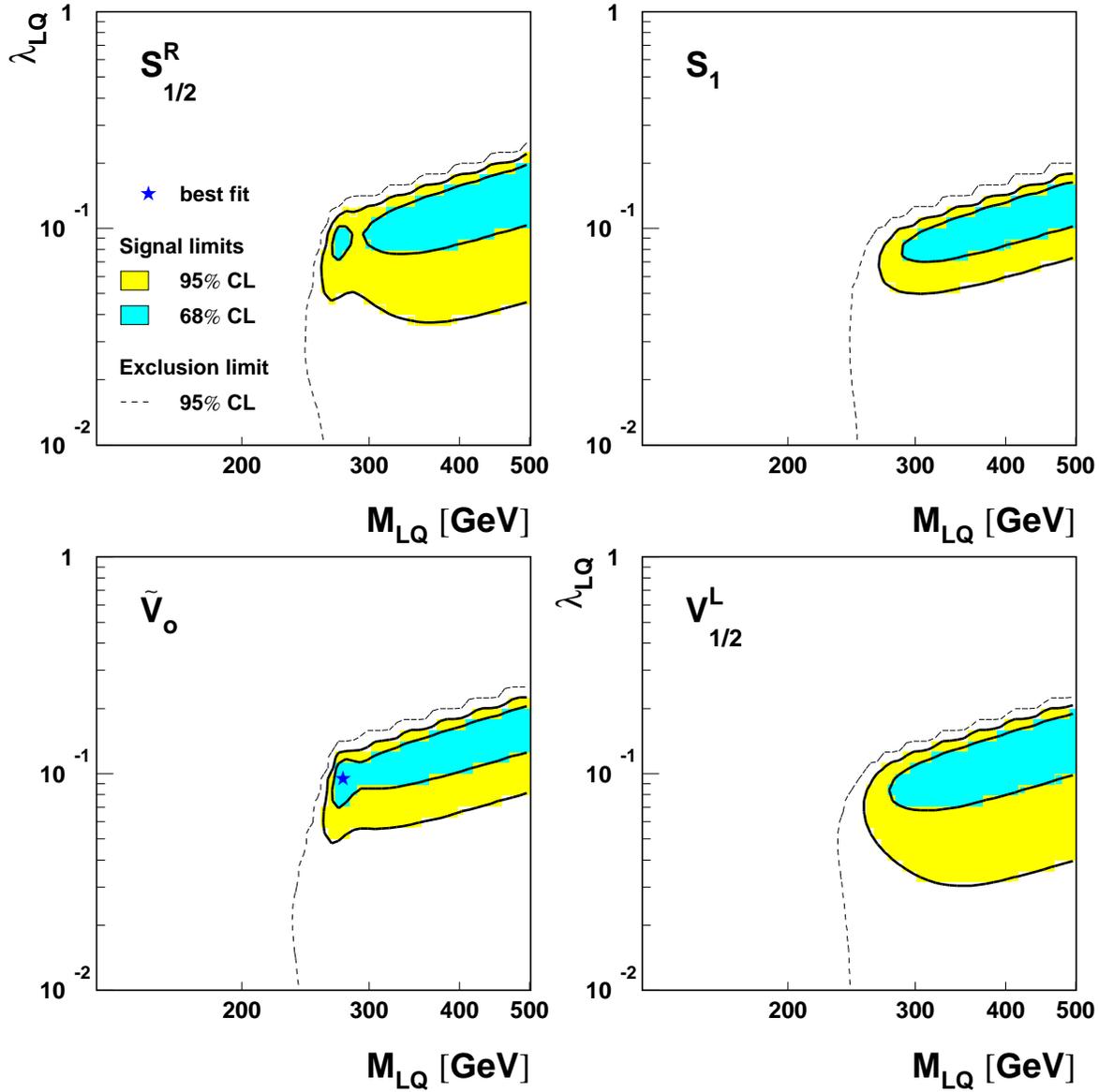}
}}
  \caption{ Signal limits on 68\% and 95\% CL for different leptoquark
            models (as indicated in the plot) resulting from the
            global analysis of existing data \cite{myglqa}.
            Dashed lines indicate the 95\% CL exclusion limits. 
            For $\tilde{V}_{\circ}$ model
            a star indicates the best fit parameters.
            For other models the best fit is obtained in the contact
            interaction limit $M_{LQ} \rightarrow \infty$.}
  \label{fig-lqsig}
\end{figure}

Four leptoquark models are found to describe the existing experimental data 
much better than the Standard Model 
(${\cal P}(\lambda_{LQ},M_{LQ})>20$).
The signal limits for these models, 
at 68\% and 95\% CL are compared with exclusion limits  
in the $(\lambda_{LQ},M_{LQ})$ space in Figure \ref{fig-lqsig}.
For $S_{1}$ and $\tilde{V}_{\circ}$ leptoquarks
the observed increase in the global probability 
by factor 367 and 142 respectively
corresponds to more than a 3$\sigma$ effect.
The leptoquark ``signal'' is mostly resulting from the new data 
on the atomic parity violation (APV) in cesium\cite{apvnew}.
After the theoretical uncertainties have been significantly reduced,
the measured value of the cesium weak charge is 
now 2.5$\sigma$ away from the Standard Model prediction.
Also the new hadronic cross-section measurements at LEP2,
for $\sqrt{s}$=192--202 GeV, are on average about 2.5\% above
the predictions\cite{lepnew}.
The effect is furthermore supported by the slight violation of
the CKM matrix unitarity and  HERA high-$Q^{2}$ results.


\section{Future experiments}
\label{sec-exp}

In the presented analysis experiments at the following existing
and future colliders are considered:
\begin{itemize}

 \item HERA \\
Since 1998 HERA collides 920 GeV protons with 27.5 GeV electrons or
positrons, resulting in $\sqrt{s} \approx$ 318 GeV. After the accelerator
upgrade in 2000/2001 HERA is expected to deliver about 200 $pb^{-1}$
of data every year. In this analysis the integrated luminosity
of  400 $pb^{-1}$ for each beam ($e^{-}p$ and $e^{+}p$) is assumed.

 \item The Tevatron \\
The Tevatron is expected to start collecting data again in 2001.
After the accelerator upgrade the $p\bar{p}$ center-of-mass energy
of $\sqrt{s}$= 2 TeV is expected and the luminosity of about 1-2 $fb^{-1}$ 
per year. Up to  10 $fb^{-1}$ of the data can be collected before 
LHC turns on. Results given in this paper were calculated for integrated 
luminosities of 1 and 10 $fb^{-1}$.

 \item LHC \\
The Large Hadron Collider (LHC), currently under construction at CERN,
will collide proton beams at the center-of-mass energy
of $\sqrt{s}$= 14 TeV. The luminosity expected at the very beginning is
about  10 $fb^{-1}$ per year and should increase up to about 
100 $fb^{-1}$ per year in the next years.
Results given in this paper were calculated for integrated 
luminosities of 10 and 100 $fb^{-1}$.

\item TESLA \\
TESLA is one of the existing proposals for the next-generation
$e^{+}e^{-}$ linear collider. It would collide 250 GeV electron
and positron beams ($\sqrt{s}$= 0.5 TeV), delivering the
integrated luminosity of up to  500 $fb^{-1}$. 
After the accelerator upgrade, beam energies of up to 500 GeV 
($\sqrt{s}$= 1 TeV) should be reachable.

High energy $e^{+}e^{-}$ collisions can be also used
to study $e\gamma$ and $\gamma \gamma$ interaction, with
the effective photon flux described by the 
Waizs{\"a}cker-Williams Approximation (WWA). 
However, high quality electron beams of TESLA could be also used to produce
high energy and high intensity photon beams from the Compton 
backscattering of laser light \cite{compton}.
In that case scattered photons usually take most of the electron energy.
The energy spectrum is much harder than for WWA and
peaked at the maximum photon energy
$E^{max}_{\gamma} \approx 0.83 \; E_{e}$.
Taking into account possible production of high intensity photon beams,
three different scenarios are considered for TESLA,
for $\sqrt{s_{ee}}$= 0.5  and 1 TeV, and the integrated luminosity 
of 100 $fb^{-1}$:
\begin{itemize}

\item $e^{+}e^{-}$ scattering; \\
  $e \gamma$  and  $\gamma \gamma$ collisions are also considered
  using the WWA effective photon flux, 

\item $e \gamma$ scattering, 
  with photon beam produced by Compton backscattering; \\
  the maximum  $e\gamma$ center-of-mass energy  
       $\sqrt{s_{e\gamma}} \approx 0.91 \sqrt{s_{ee}}$,

\item $\gamma \gamma$ scattering, 
  with both photon beams produced by Compton backscattering; \\
  the maximum  $\gamma\gamma$ center-of-mass energy  
       $\sqrt{s_{\gamma\gamma}} \approx 0.83 \sqrt{s_{ee}}$,

\end{itemize}

\item THERA \\
If TESLA project is approved, it is also possible to consider
scattering of 250 or 500~GeV electron beam from TESLA with 
1 TeV proton beam from HERA, resulting in the center-of-mass energy
of $\sqrt{s}$= 1 and 1.4 TeV respectively. Expected integrated luminosity
is of the order of 100 $pb^{-1}$.

\end{itemize}


\section{Contact interaction limit}
\label{sec-ci}

In the limit $M_{LQ} \gg \sqrt{s}$  the effect of leptoquark 
production or exchange is equivalent to a vector type $eeqq$ 
contact interaction \cite{lqci}.
Limits on the effective contact interaction mass scale $\Lambda$
(related to the leptoquark mass to the coupling ratio $M_{LQ}/\lambda_{LQ}$)
can be extracted from precise measurements of different
Standard Model processes.
Expected limits from future experiments are calculated assuming that
no deviations from the Standard Model predictions will be observed.
The method used has been described in details in \cite{myglqa,mygcia}.
Limits from the following future measurements are considered:
\begin{itemize}
\item Drell-Yan electron pair production at the Tevatron and LHC; \\
      from comparison with \cite{cdfdy,cmsdy} event selection efficiency 
      is assumed to be 50\% at the Tevatron  and 80\% at LHC,
\item hadronic cross-section measurements at TESLA $e^{+}e^{-}$, \\
     assuming that $\sigma(e^{+}e^{-} \rightarrow q \bar{q})$ 
     is measured with 1\% precision,
\item high-$Q^{2}$ NC DIS cross-section for $e^{\pm}p$ scattering at HERA
      and $e^{-}p$ scattering at THERA,
\item high-$Q^{2}$ NC DIS cross-section for $e\gamma$ scattering at TESLA, 
     for $e^{+}e^{-}$ ($\gamma$ from WWA) and $e\gamma$ ($\gamma$ from 
     Compton backscattering) scenarios. 
\end{itemize}
Parton densities are described by the MRST distribution functions \cite{mrst}
for the proton and GRV LO distribution functions \cite{grv} for the photon.
\begin{table}[tbp]
\begin{center}
\begin{tabular}{lccccccc}
\hline\hline\hline\noalign{\smallskip}
 & Current & \multicolumn{6}{c}%
{Expected 95\% CL exclusion limits on $M_{LQ}/\lambda_{LQ}$ [TeV]} \\
\cline{3-8}\noalign{\smallskip}
Model  & limits on & \multicolumn{2}{c}{$p \bar{p}$}       &  
                  \multicolumn{2}{c}{$pp$} &
                  \multicolumn{2}{c}{$e^{+}e^{-}$} \\
   &  $M_{LQ}/\lambda_{LQ}$  & \multicolumn{2}{c}{$\sqrt{s}$=2 TeV} & 
                  \multicolumn{2}{c}{$\sqrt{s}$=14 TeV} &
                  \multicolumn{2}{c}{$\Delta \sigma_{had}$ = 1\% } \\
  & [TeV] &  1 $fb^{-1}$  & 10 $fb^{-1}$  &   10 $fb^{-1}$  &  100 $fb^{-1}$  &
                                  $\sqrt{s}$=0.5 TeV &   1 TeV \\
\hline\noalign{\smallskip}
$S_{\circ}^{L}     $
 & 3.7 & 1.3 & 2.1 & 4.5 & 7.5 & 3.5 & 6.7  \\
$S_{\circ}^{R}     $
 & 3.9 & 0.9 & 1.6 & 3.4 & 5.8 & 3.0 & 5.6  \\
$\tilde{S}_{\circ} $
 & 3.6 & 0.6 & 0.9 & 3.3 & 4.9 & 2.2 & 4.5  \\
$S_{1/2}^{L}       $
 & 3.5 & 1.1 & 1.6 & 4.0 & 6.0 & 2.2 & 4.5  \\
$S_{1/2}^{R}       $
 & 2.1 & 1.0 & 1.5 & 4.2 & 6.3 & 2.4 & 4.9  \\
$\tilde{S}_{1/2}   $
 & 3.8 & 0.6 & 0.8 & 2.8 & 4.0 & 1.3 & 2.6  \\
$S_{1}             $
 & 2.4 & 0.8 & 1.2 & 4.6 & 6.4 & 3.1 & 6.3  \\ 
\hline\noalign{\smallskip}
$V_{\circ}^{L}     $
 & 8.1 & 0.7 & 1.0 & 4.1 & 6.2 & 4.5 & 8.5  \\
$V_{\circ}^{R}     $
 & 2.3 & 0.8 & 1.1 & 3.9 & 5.4 & 1.6 & 3.1  \\
$\tilde{V}_{\circ} $
 & 1.9 & 1.8 & 2.7 & 6.6 & 10.1 & 4.0 & 8.4  \\
$V_{1/2}^{L}       $
 & 2.1 & 0.9 & 1.2 & 4.5 & 6.4 & 2.6 & 5.4  \\
$V_{1/2}^{R}       $
 & 7.5 & 1.2 & 1.8 & 5.1 & 7.2 & 2.0 & 4.0  \\
$\tilde{V}_{1/2}   $
 & 2.1 & 1.2 & 1.6 & 4.4 & 6.3 & 1.6 & 3.1  \\
$V_{1}             $
 & 7.3 & 2.7 & 4.2 & 9.5 & 14.9 & 5.2 & 11.0  \\

\hline\hline\hline\noalign{\smallskip}
\end{tabular}
\end{center}
\caption{95\% CL exclusion limits on $M_{LQ}/\lambda_{LQ}$, expected from 
         future measurements of the Drell-Yan electron pair production at
         the Tevatron and LHC, and the total hadronic cross-section
         measurements at TESLA. Current experimental limits
         are included for comparison.
         \label{tab-ci}}
\end{table}
95\% CL exclusion limits on $M_{LQ}/\lambda_{LQ}$, expected from 
different future experiments, are presented in Tables \ref{tab-ci} and 
\ref{tab-ci2}.
Current limits from global analysis \cite{myglqa} are included for comparison.
Strongest limits on the leptoquark mass to the coupling ratio,
in the contact interaction approximation are expected from
the Drell-Yan electron pair production at  LHC and
hadronic cross-section measurements at TESLA.
Results presented 
for Drell-Yan electron pair production and high-$Q^{2}$ NC DIS measurements
are the mean values from about 2000 MC experiments.
Poisson fluctuations in the observed numbers of events can result in
the statistical fluctuation of the actual limit of 10-20\%.  
\begin{table}[tbp]
\begin{center}
\begin{tabular}{lcccccccc}
\hline\hline\hline\noalign{\smallskip}
 & Current & \multicolumn{7}{c}%
{Expected 95\% CL exclusion limits on $M_{LQ}/\lambda_{LQ}$ [TeV]} \\
\cline{3-9}\noalign{\smallskip}
Model  & limits on & $e^{\pm}p$  & 
                  \multicolumn{4}{c}{$e^{+}e^{-}$   100 $fb^{-1}$}       &  
                  \multicolumn{2}{c}{$e^{-}p$} \\
   & $M_{LQ}/\lambda_{LQ}$ & $\sqrt{s}$=318GeV &
               \multicolumn{2}{c}{$\sqrt{s_{ee}}$=0.5 TeV} & 
                \multicolumn{2}{c}{$\sqrt{s_{ee}}$=1 TeV} &
                \multicolumn{2}{c}{100 $pb^{-1}$} \\
   & [TeV]  &  2$\times$400 $pb^{-1}$  & 
           $e^{+}e^{-}$ & $e^{-}\gamma$ &  $e^{+}e^{-}$ & $e^{-}\gamma$ &
                       $\sqrt{s}$=1 TeV &   1.4 TeV \\
\hline\noalign{\smallskip}
$S_{\circ}^{L}     $
 & 3.7 & 1.0 & 1.1 & 1.4 & 1.8 & 2.5 & 1.7 & 2.0  \\
$S_{\circ}^{R}     $
 & 3.9 & 0.9 & 1.0 & 1.3 & 1.6 & 2.3 & 1.5 & 1.8  \\
$\tilde{S}_{\circ} $
 & 3.6 & 0.4 & 0.4 & 0.5 & 0.7 & 0.9 & 0.7 & 0.9  \\
$S_{1/2}^{L}       $
 & 3.5 & 0.9 & 0.6 & 0.7 & 1.0 & 1.2 & 0.6 & 0.8  \\
$S_{1/2}^{R}       $
 & 2.1 & 0.6 & 0.6 & 0.8 & 1.1 & 1.4 & 0.7 & 0.9  \\
$\tilde{S}_{1/2}   $
 & 3.8 & 0.5 & 0.5 & 0.6 & 0.8 & 1.1 & 0.6 & 0.8  \\
$S_{1}             $
 & 2.4 & 0.8 & 1.0 & 1.3 & 1.6 & 2.3 & 1.4 & 1.7  \\
\hline\noalign{\smallskip}
$V_{\circ}^{L}     $
 & 8.1 & 0.9 & 0.8 & 1.1 & 1.4 & 1.9 & 1.6 & 2.0  \\
$V_{\circ}^{R}     $
 & 2.3 & 0.7 & 0.7 & 0.9 & 1.2 & 1.7 & 1.3 & 1.7  \\
$\tilde{V}_{\circ} $
 & 1.9 & 1.2 & 1.1 & 1.5 & 1.7 & 2.6 & 1.8 & 2.2  \\
$V_{1/2}^{L}       $
 & 2.1 & 0.6 & 0.6 & 0.8 & 1.1 & 1.4 & 0.8 & 1.0  \\
$V_{1/2}^{R}       $
 & 7.5 & 1.3 & 1.1 & 1.4 & 1.9 & 2.5 & 1.2 & 1.5  \\
$\tilde{V}_{1/2}   $
 & 2.1 & 1.4 & 1.1 & 1.5 & 1.9 & 2.6 & 1.1 & 1.4  \\
$V_{1}             $
 & 7.3 & 1.9 & 1.8 & 2.4 & 2.8 & 4.3 & 2.6 & 3.0  \\
\hline\hline\hline\noalign{\smallskip}
\end{tabular}
\end{center}
\caption{95\% CL exclusion limits on $M_{LQ}/\lambda_{LQ}$, expected from 
         future measurements of high-$Q^{2}$ NC DIS cross-sections 
         at HERA ($e^{\pm}p$ scattering), TESLA ($e\gamma$ scattering 
         in   $e^{+}e^{-}$ and $e\gamma$ scenarios) and THERA ($e^{-}p$). 
         Current experimental limits  are included for comparison.
                  \label{tab-ci2}}
\end{table}


\section{Limits from leptoquark pair production}
\label{sec-pair}

Leptoquark pair production has been considered for $p\bar{p}$
collisions at the Tevatron,  $pp$ collisions at LHC and
$\gamma\gamma$ scattering at TESLA (in $e^{+}e^{-}$ and $\gamma\gamma$
scenarios). 
The advantage of the leptoquark pair production is that
the cross-section depends only on the the strong
or electromagnetic coupling constant and does not depend on the leptoquark 
Yukawa coupling.
The leptoquark mass limits derived from the search for the
leptoquark pair production are therefor valid for arbitrary values 
of $\lambda_{LQ}$.

Presented results are based on the cross-sections given in 
\cite{jb1,jb2}.
For vector leptoquark production at the Tevatron and LHC the
anomalous coupling values resulting in the minimal pair production
cross-section are assumed.
For vector leptoquark production at TESLA minimal couplings 
$\kappa$=1, $\lambda$=0 are used.
For leptoquark pair production at LHC the Standard Model background
estimate is taken from \cite{cmslq}.
For leptoquark pair production searches at the Tevatron and at TESLA the
Standard Model background is assumed to be negligible.
Event selection efficiency is 25\% at the Tevatron \cite{darin}
and 30\% at LHC (from comparison with \cite{cmslq}).
For TESLA it is assumed that (due to much ``cleaner'' environment
of $e^{+}e^{-}$ collisions) high event selection efficiency
(close to 100\%) is possible.
95\% CL exclusion limits on the leptoquark mass $M_{LQ}$, expected from 
the negative search results at different future experiments, 
are presented in Table \ref{tab-lq}.
In all cases strongest limits on the leptoquark mass  are expected from
the pair production search at  LHC.

\begin{table}[tb]
\begin{center}
\begin{tabular}{lccccccccc}
\hline\hline\hline\noalign{\smallskip}
 & Current & 
 \multicolumn{8}{c}{Expected 95\% CL exclusion limits on $M_{LQ}$ [GeV]} \\
\cline{3-10}\noalign{\smallskip}
 Model & limits  & \multicolumn{2}{c}{$p \bar{p}$}  &  
                  \multicolumn{2}{c}{$pp$} &
    \multicolumn{4}{c}{$e^{+}e^{-}$  ~~~100 $fb^{-1}$ } \\
 & on $M_{LQ}$  & \multicolumn{2}{c}{$\sqrt{s}$=2 TeV} & 
                  \multicolumn{2}{c}{$\sqrt{s}$=14 TeV} &
                  \multicolumn{2}{c}{$\sqrt{s_{ee}}$=0.5 TeV} &
                  \multicolumn{2}{c}{$\sqrt{s_{ee}}$=1 TeV} \\
 & [GeV]  & 1 $fb^{-1}$ & 10 $fb^{-1}$   &   10 $fb^{-1}$  &  100 $fb^{-1}$  &
       $e^{+}e^{-}$ &  $\gamma \gamma$ & $e^{+}e^{-}$ &  $\gamma \gamma$ \\
\hline\noalign{\smallskip}
$S_{\circ}^{L}     $
 & 213 & 255 & 335 & 870 & 1160 & 151 & 197 & 259 & 376  \\
$S_{\circ}^{R}     $
 & 242 & 300 & 380 & 1170 & 1520 & 168 & 202 & 296 & 395  \\
$\tilde{S}_{\circ} $
 & 242 & 300 & 380 & 1170 & 1520 & 222 & 207 & 415 & 412  \\
$S_{1/2}^{L}       $
 & 229 & 300 & 380 & 1170 & 1520 & 229 & 207 & 437 & 413  \\
$S_{1/2}^{R}       $
 & 245 & 325 & 405 & 1310 & 1700 & 230 & 207 & 437 & 413  \\
$\tilde{S}_{1/2}   $
 & 233 & 300 & 380 & 1170 & 1520 & 188 & 206 & 332 & 408  \\
$S_{1}             $
 & 245 & 310 & 390 & 1220 & 1580 & 222 & 207 & 415 & 412  \\
\hline\noalign{\smallskip}
$V_{\circ}^{L}     $
 & 230 & 275 & 335 & 1270 & 1630 & 189 & 205 & 343 & 408  \\
$V_{\circ}^{R}     $
 & 231 & 310 & 375 & 1540 & 1990 & 204 & 206 & 379 & 411  \\
$\tilde{V}_{\circ} $
 & 235 & 310 & 375 & 1540 & 1990 & 236 & 207 & 456 & 414  \\
$V_{1/2}^{L}       $
 & 235 & 310 & 375 & 1540 & 1990 & 230 & 207 & 441 & 413  \\
$V_{1/2}^{R}       $
 & 262 & 330 & 395 & 1670 & 2160 & 230 & 207 & 442 & 413  \\
$\tilde{V}_{1/2}   $
 & 244 & 310 & 375 & 1540 & 1990 & 188 & 204 & 346 & 403  \\
$V_{1}             $
 & 254 & 315 & 385 & 1580 & 2040 & 236 & 207 & 456 & 414  \\
\hline\hline\hline\noalign{\smallskip}
\end{tabular}
\end{center}
\caption{95\% CL exclusion limits on the leptoquark mass $M_{LQ}$, 
         expected from the negative pair production search results 
         at the Tevatron, LHC and TESLA.
         Current experimental limits  are included for comparison.
         \label{tab-lq}}
\end{table}


\section{Limits from single leptoquark production}
\label{sec-single}

Production of single leptoquarks has been considered for 
$e^{\pm}p$ scattering at HERA and THERA and for
$e\gamma$ scattering at TESLA (in $e^{+}e^{-}$ and $e\gamma$
scenarios). 
In the narrow-width approximation, the cross-section for single 
$F=2$ leptoquark production in electron-proton scattering 
(via the electron-quark fusion) is given by:
\begin{eqnarray}
\sigma^{ep\rightarrow LQ \; X}(M_{LQ},\lambda_{LQ}) & = & 
(J+1) \cdot \frac{\pi \lambda_{LQ}^{2}}{4 M_{LQ}^{2}} 
        \cdot x_{LQ} q(x_{LQ},M_{LQ}^{2}) \label{eq-scs}
\end{eqnarray}
where $J$ is the leptoquark spin,
$q(x,Q^{2})$ is the quark momentum distribution function 
in the proton and $x_{LQ} =\frac{M_{LQ}^{2}}{s}$.
For the single leptoquark production in $e\gamma$ collisions
two approaches are possible:
\begin{itemize}
  \item leptoquark is produced in the electron fusion
        with a quark inside the photon.
        The production cross-section is given by
        the formula \eqref{eq-scs}, with $q(x,Q^{2})$ describing
        the quark momentum distribution in the photon.
        As before, GRV LO parton densities for the photon\cite{grv} are used.
  \item photon directly participates in the
        process $e\gamma \rightarrow LQ \; q$. The cross-section
        for this process is taken from \cite{egdir}.
\end{itemize}
Both approaches give very similar numerical results for the
single leptoquark production cross-section.
This is due to the fact that
the dominant contribution to the direct photon
process comes from the diagram in which photon splits into the 
$q\bar{q}$ pair,
which is also described in the resolved photon approach,
mentioned above.

\begin{figure}[tbp]
\centerline{\resizebox{\figwidth}{!}{%
  \includegraphics{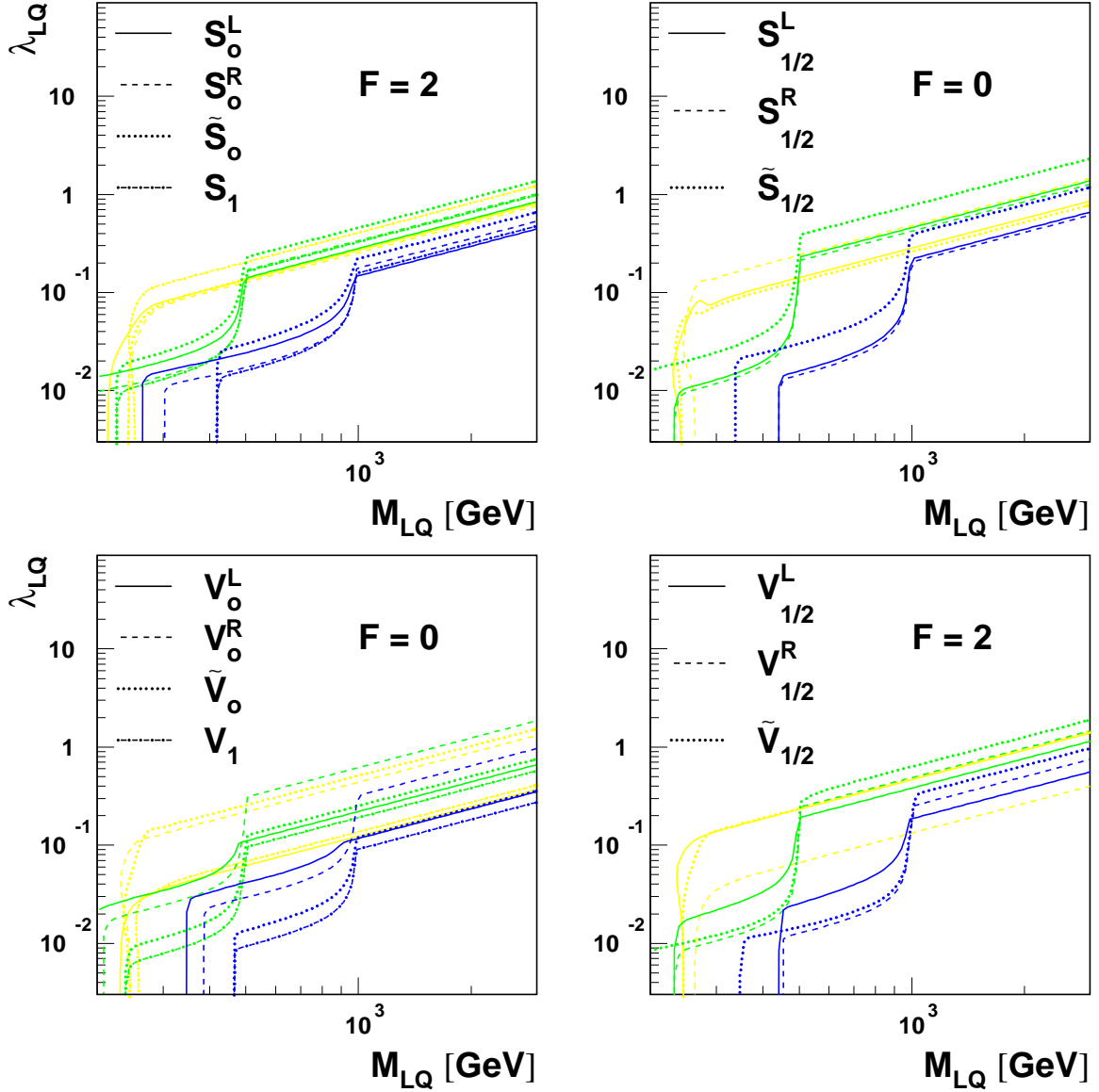}}}
  \caption{ Expected 95\% CL exclusion limits in  $(\lambda_{LQ},M_{LQ})$
            space, for different leptoquark  models (as indicated in the plot),
            for 100 $fb^{-1}$ of $e^{+}e^{-}$ data at TESLA.
            Combined limits (based on leptoquark pair production, 
            single leptoquark production and hadronic cross-section
            measurements) for electron beam energy of 500 GeV (dark blue) 
           and 250 GeV (light green) are compared with existing limits 
           (yellow; very light).
           }
  \label{fig-ee}
\end{figure}

\begin{figure}[tbp]
\centerline{\resizebox{\figwidth}{!}{%
  \includegraphics{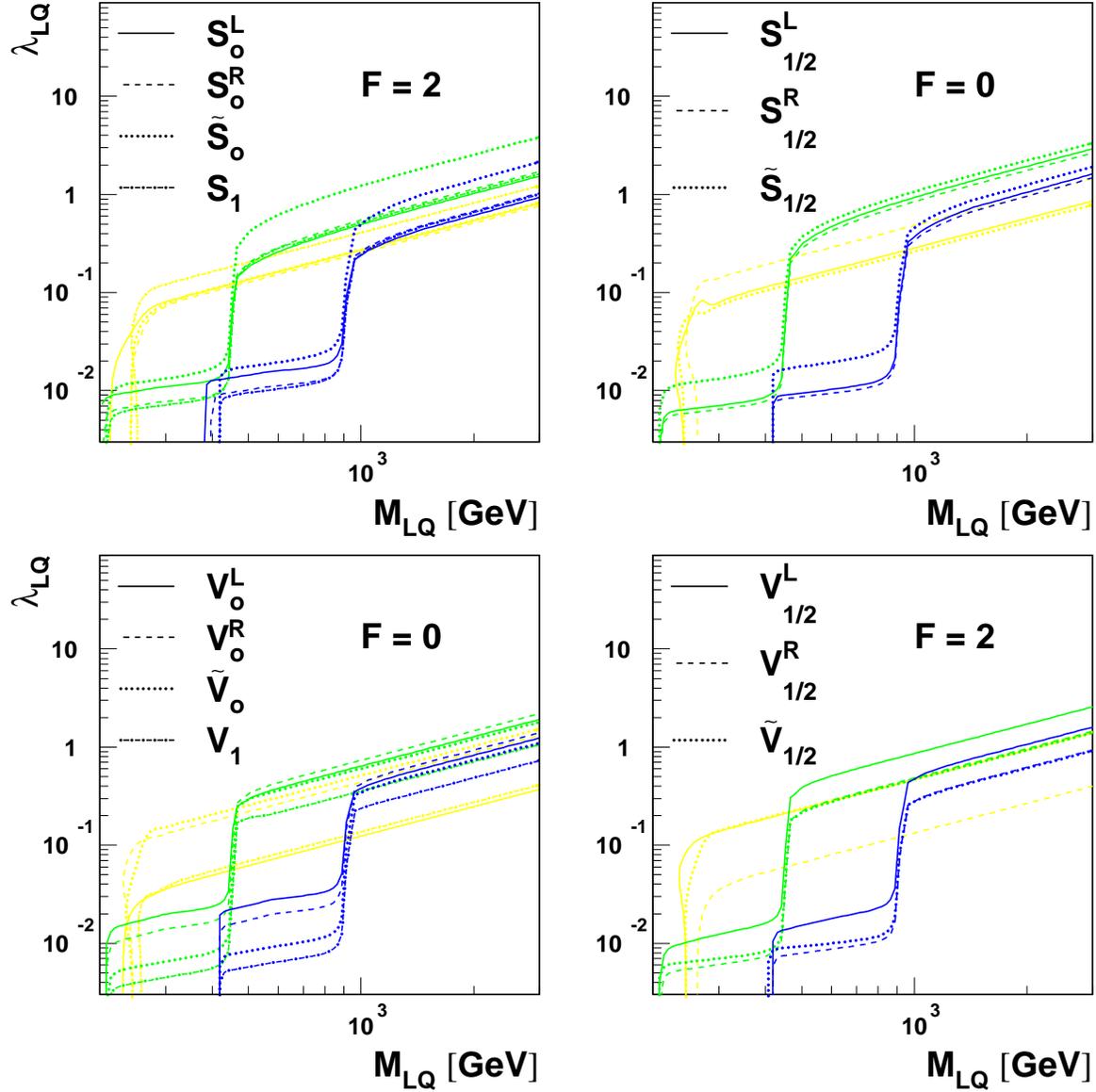}}}
  \caption{ Expected 95\% CL exclusion limits in  $(\lambda_{LQ},M_{LQ})$
            space, for different leptoquark  models (as indicated in the plot),
            for 100 $fb^{-1}$ of $e\gamma$ and $\gamma\gamma$ data at TESLA.
            Combined limits (based on leptoquark pair production in 
            $\gamma\gamma$, single leptoquark production in $e\gamma$ 
            and $e\gamma$ NC DIS cross-section measurements) 
            for electron beam energy of 500 GeV (dark blue) 
           and 250 GeV (light green) are compared with existing limits 
           (yellow; very light).
           }
  \label{fig-gg}
\end{figure}

\begin{figure}[tbp]
\centerline{\resizebox{\figwidth}{!}{%
  \includegraphics{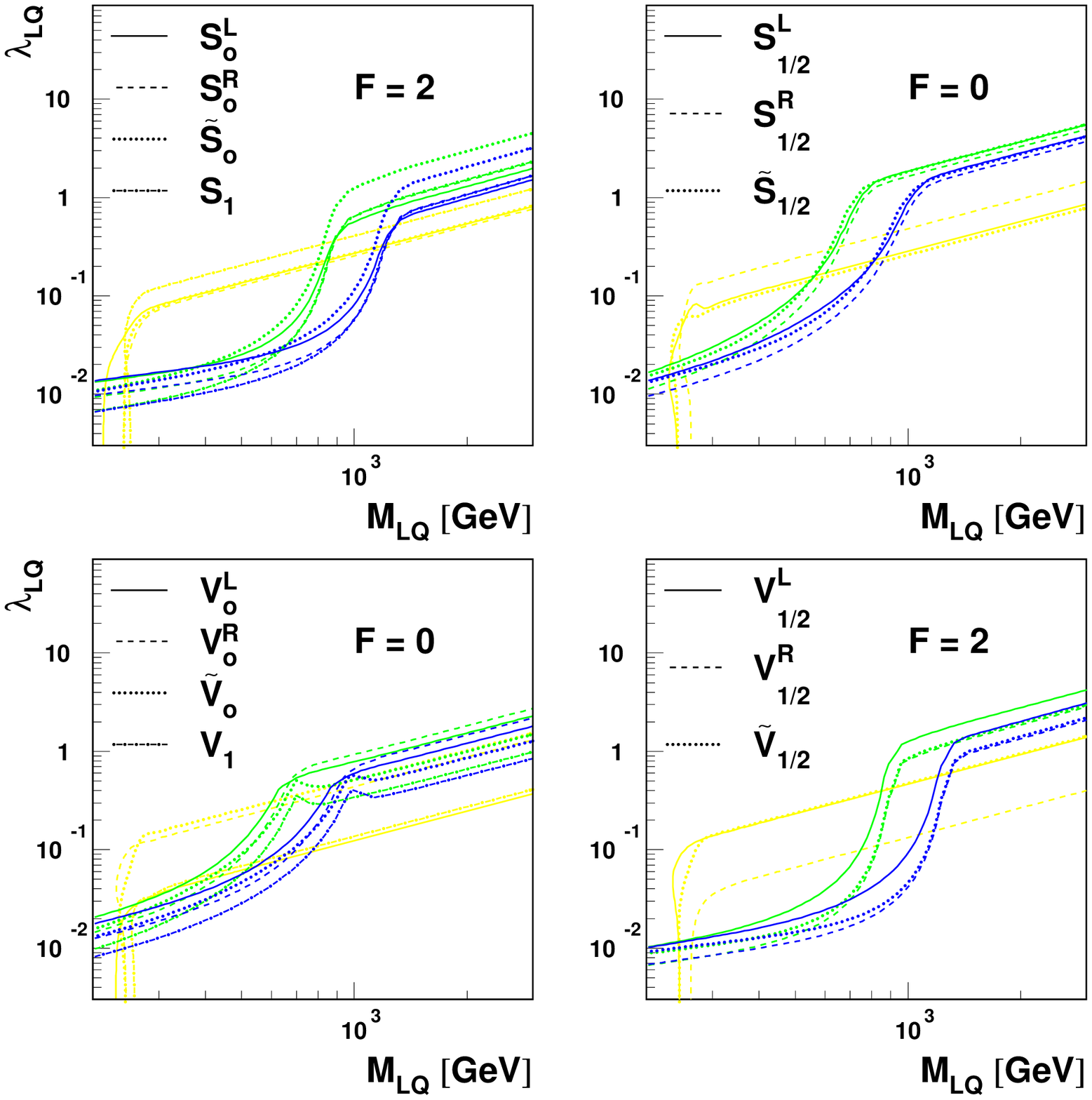}
}}
  \caption{ Expected 95\% CL exclusion limits in  $(\lambda_{LQ},M_{LQ})$
            space, for different leptoquark  models (as indicated in the plot),
            for 100 $pb^{-1}$ of $e^{-}p$ data at THERA.
            Combined limits (based on single leptoquark production 
            and high-$Q^{2}$ NC DIS cross-section  measurements) 
            for electron beam energy of 500 GeV (dark blue) 
           and 250 GeV (light green) are compared with existing limits 
           (yellow; very light).
           }
  \label{fig-tera}
\end{figure}

Only the leptoquark signal in the electron-jet decay channel is considered
in this study and the resolution of the mass reconstruction is 
assumed to be 5\%.
Expected signal from single leptoquark production,
for given leptoquark mass $M_{LQ}$ and Yukawa coupling $\lambda_{LQ}$, 
is compared with the observed number of events from the Standard Model 
background ($ep$ or $e\gamma$ NC DIS).
The background can be suppressed by applying a cut on 
the Bjorken variable $y$, which is optimized for every leptoquark 
type as a function of the leptoquark mass.
The 95\%CL exclusion limit on the leptoquark Yukawa coupling
$\lambda_{LQ}$ corresponds to the decrease of the Poisson 
probability for the number of events observed in the leptoquark 
mass window to 5\% of the Standard Model probability.
Average $\lambda_{LQ}$ exclusion limits expected from
future experiments are calculated based on about 2000 MC 
experiments generated according to the Standard Model expectations.
In Figures \ref{fig-ee}, \ref{fig-gg} and \ref{fig-tera}
combined 95\% CL exclusion limits in  $(\lambda_{LQ},M_{LQ})$
space, expected from single leptoquark production,
leptoquark pair production and contact interaction analysis,
are presented for future $e^{+}e^{-}$ scattering at TESLA,
$e\gamma$ and $\gamma\gamma$ scattering at TESLA and
$ep$ scattering at THERA, respectively.
Results obtained for TESLA electron beam energy of 250 and 500 GeV
are compared with existing limits.
In all cases search for single leptoquark production significantly 
improves the limits coming from leptoquark pair production and
indirect searches.
For coupling values  $\lambda_{LQ}\sim$ 0.1 the leptoquark mass
limits can be extended up to the kinematic limit $M_{LQ}) \sim \sqrt{s}$.
For lower leptoquark masses (but above the limit from the leptoquark
pair production), limits on the leptoquark Yukawa coupling $\lambda_{LQ}$ 
can be improved by an order of magnitude.


\section{Summary of results}
\label{sec-comp}

\begin{figure}[tbp]
\centerline{\resizebox{\figwidth}{!}{%
  \includegraphics{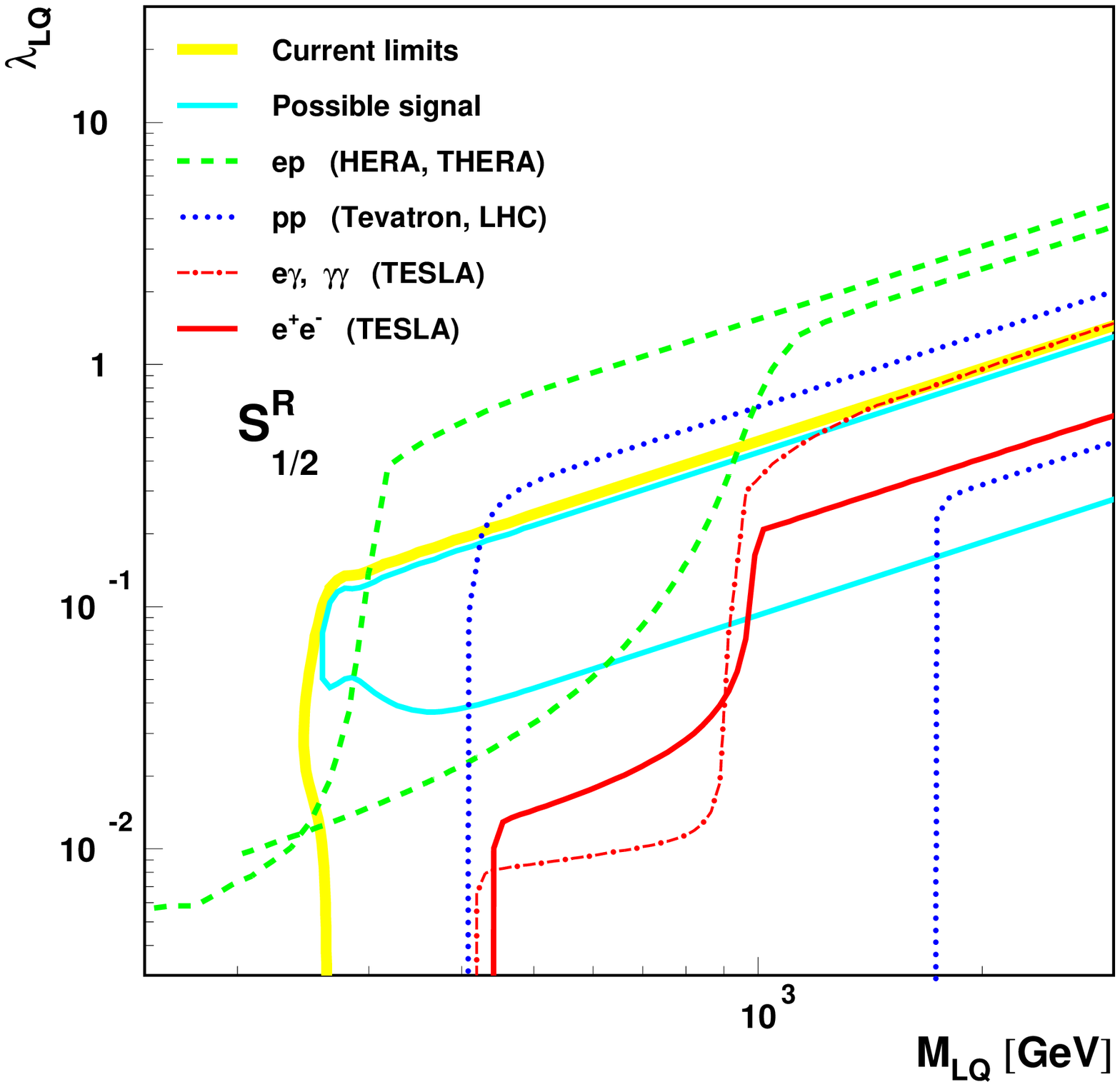}
}}
  \caption{ Comparison of expected 95\% CL exclusion limits 
            in  $(\lambda_{LQ},M_{LQ})$ for $S_{1/2}^R$ leptoquark  model,
            for different future experiments, as indicated in the plot.
            Presented limits correspond to 
2$\times$400 $pb^{-1}$ of $e^{\pm}p$ data at HERA ($\sqrt{s}$=318 GeV),
100 $pb^{-1}$ of $e^{-}p$ data at THERA ($\sqrt{s}$=1.4 TeV),
10 $fb^{-1}$ of $p\bar{p}$ data at the Tevatron ($\sqrt{s}$=2 TeV),
100 $fb^{-1}$ of $p\bar{p}$ data at the LHC ($\sqrt{s}$=14 TeV)
and 100 $fb^{-1}$ of $e^{+}e^{-}$, $e\gamma$ and $\gamma\gamma$ data 
      at TESLA ($\sqrt{s_{ee}}$=1 TeV).
  Also indicated are 95\% CL exclusion and signal limits from global analysis 
  of existing data\cite{myglqa}.
           }
  \label{fig-comp5}
\end{figure}

\begin{figure}[tbp]
\centerline{\resizebox{\figwidth}{!}{%
  \includegraphics{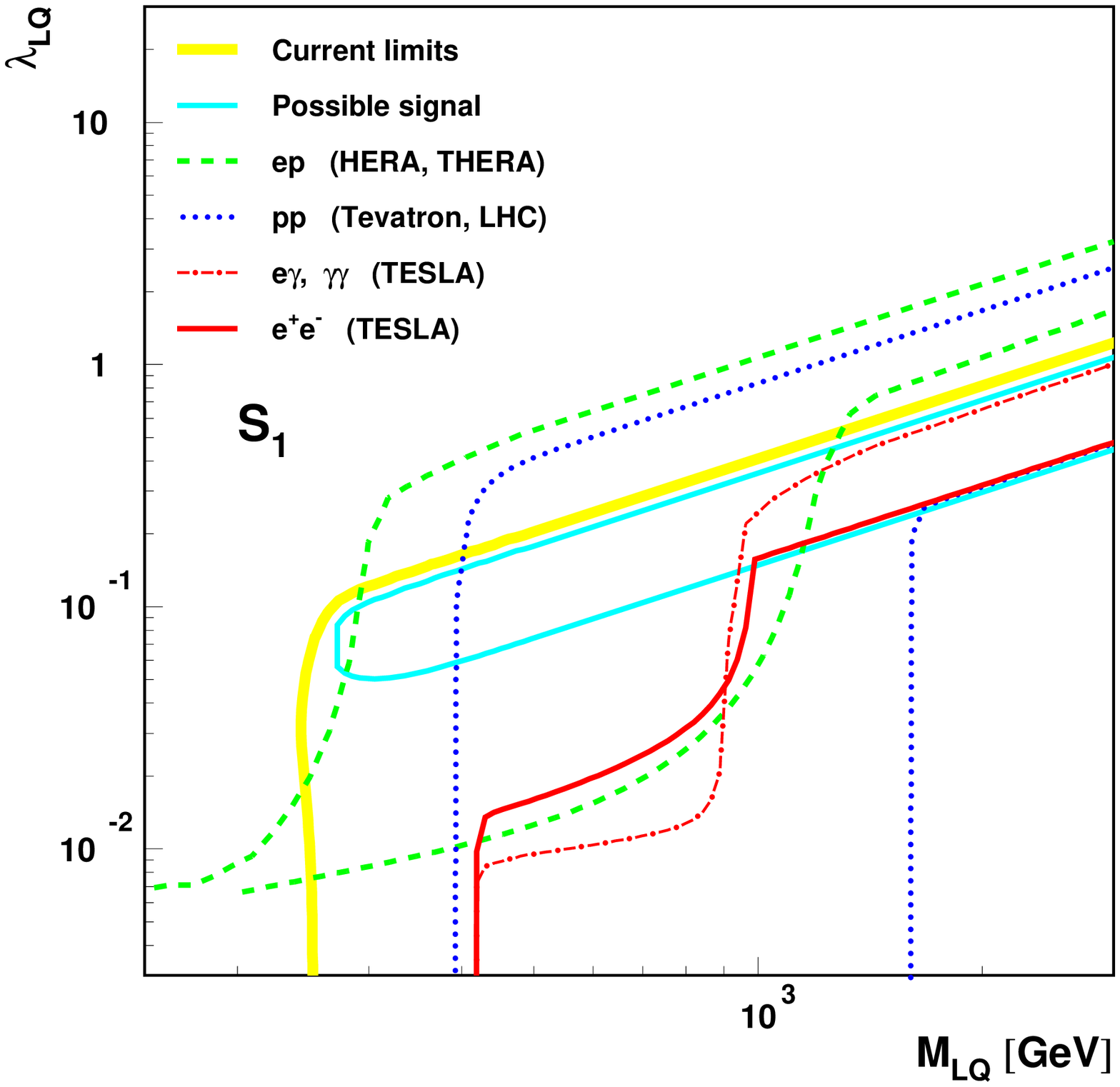}
}}
  \caption{ Comparison of expected 95\% CL exclusion limits 
            in  $(\lambda_{LQ},M_{LQ})$ for $S_{1}$ leptoquark  model,
            for different future experiments, as indicated in the plot.
            Presented limits correspond to 
2$\times$400 $pb^{-1}$ of $e^{\pm}p$ data at HERA ($\sqrt{s}$=318 GeV),
100 $pb^{-1}$ of $e^{-}p$ data at THERA ($\sqrt{s}$=1.4 TeV),
10 $fb^{-1}$ of $p\bar{p}$ data at the Tevatron ($\sqrt{s}$=2 TeV),
100 $fb^{-1}$ of $p\bar{p}$ data at the LHC ($\sqrt{s}$=14 TeV)
and 100 $fb^{-1}$ of $e^{+}e^{-}$, $e\gamma$ and $\gamma\gamma$ data 
      at TESLA ($\sqrt{s_{ee}}$=1 TeV).
  Also indicated are 95\% CL exclusion and signal limits from global analysis 
  of existing data\cite{myglqa}.
           }
  \label{fig-comp7}
\end{figure}

\begin{figure}[tbp]
\centerline{\resizebox{\figwidth}{!}{%
  \includegraphics{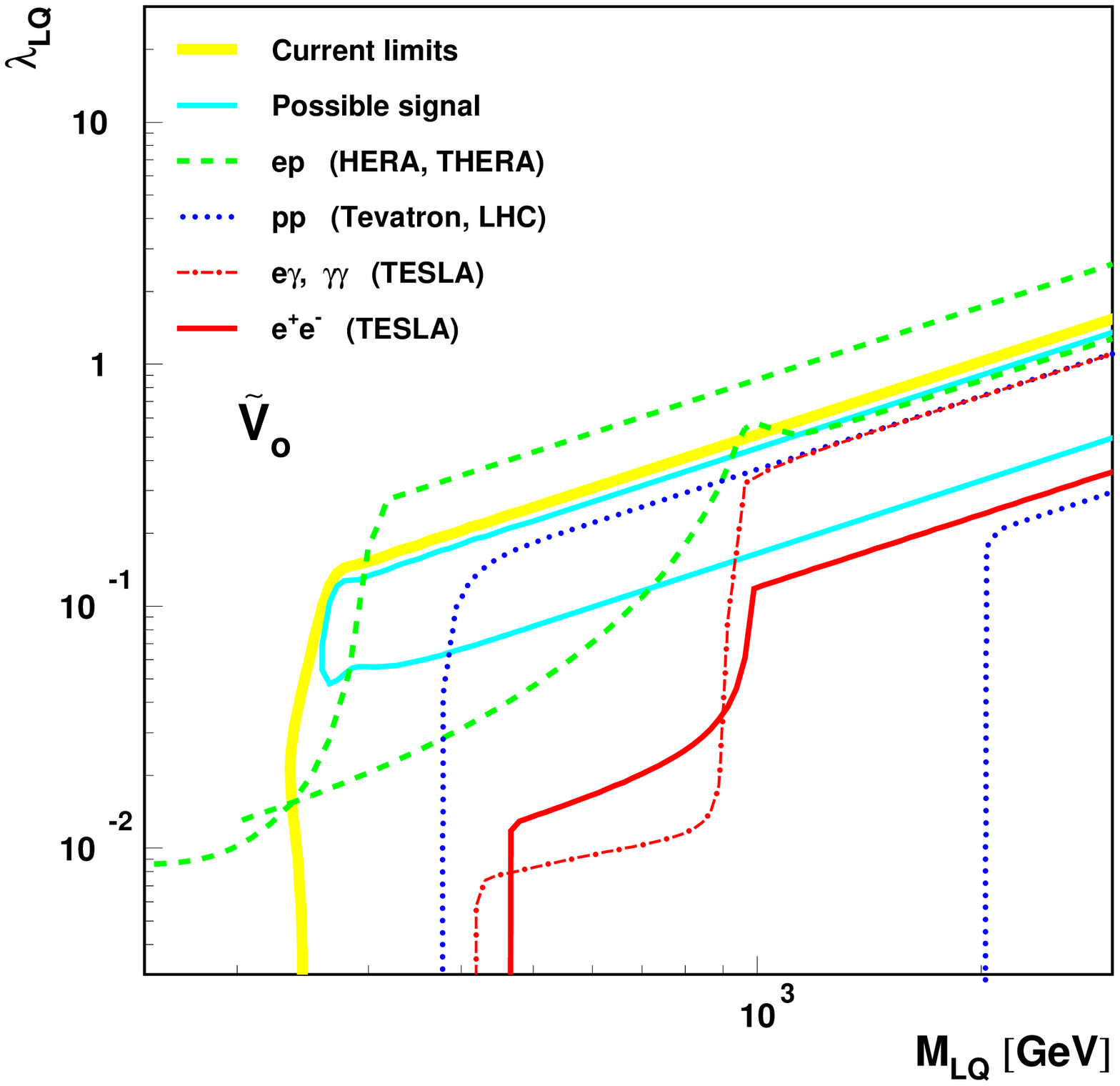}
}}
  \caption{ Comparison of expected 95\% CL exclusion limits 
       in  $(\lambda_{LQ},M_{LQ})$ for $\tilde{V}_{\circ}$ leptoquark  model,
            for different future experiments, as indicated in the plot.
            Presented limits correspond to 
2$\times$400 $pb^{-1}$ of $e^{\pm}p$ data at HERA ($\sqrt{s}$=318 GeV),
100 $pb^{-1}$ of $e^{-}p$ data at THERA ($\sqrt{s}$=1.4 TeV),
10 $fb^{-1}$ of $p\bar{p}$ data at the Tevatron ($\sqrt{s}$=2 TeV),
100 $fb^{-1}$ of $p\bar{p}$ data at the LHC ($\sqrt{s}$=14 TeV)
and 100 $fb^{-1}$ of $e^{+}e^{-}$, $e\gamma$ and $\gamma\gamma$ data 
      at TESLA ($\sqrt{s_{ee}}$=1 TeV).
  Also indicated are 95\% CL exclusion and signal limits from global analysis 
  of existing data\cite{myglqa}.
           }
  \label{fig-comp10}
\end{figure}

\begin{figure}[tbp]
\centerline{\resizebox{\figwidth}{!}{%
  \includegraphics{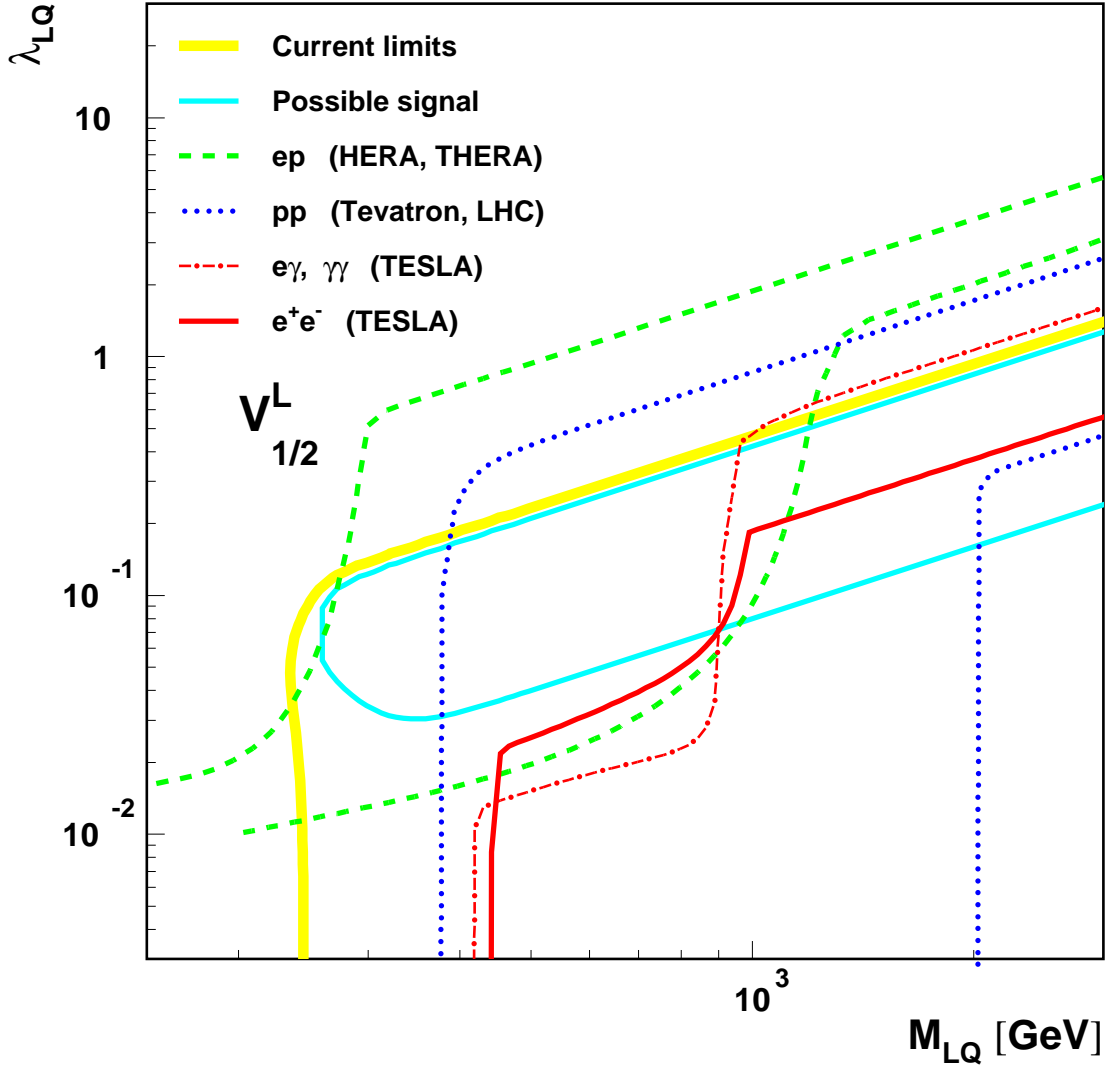}
}}
  \caption{ Comparison of expected 95\% CL exclusion limits 
            in  $(\lambda_{LQ},M_{LQ})$ for $V_{1/2}^L$ leptoquark  model,
            for different future experiments, as indicated in the plot.
            Presented limits correspond to 
2$\times$400 $pb^{-1}$ of $e^{\pm}p$ data at HERA ($\sqrt{s}$=318 GeV),
100 $pb^{-1}$ of $e^{-}p$ data at THERA ($\sqrt{s}$=1.4 TeV),
10 $fb^{-1}$ of $p\bar{p}$ data at the Tevatron ($\sqrt{s}$=2 TeV),
100 $fb^{-1}$ of $p\bar{p}$ data at the LHC ($\sqrt{s}$=14 TeV)
and 100 $fb^{-1}$ of $e^{+}e^{-}$, $e\gamma$ and $\gamma\gamma$ data 
      at TESLA ($\sqrt{s_{ee}}$=1 TeV).
  Also indicated are 95\% CL exclusion and limits from global analysis 
  of existing data\cite{myglqa}.
           }
  \label{fig-comp11}
\end{figure}

In Figures \ref{fig-comp5}, \ref{fig-comp7}, \ref{fig-comp10} and 
\ref{fig-comp11}, 95\% CL exclusion limits in  $(\lambda_{LQ},M_{LQ})$
space, expected from different experiments are compared with existing limits
and possible leptoquark signal, for $S_{1/2}^R$, $S_1$, $\tilde{V}_{\circ}$
and $V_{1/2}^L$ leptoquark models respectively. 
New HERA data are not expected to improve the existing
leptoquark limits significantly. 
Also the effect of future measurements at the Tevatron is moderate:
leptoquark mass limits will increase by 100-150 GeV (see also
Table \ref{tab-lq}).
Sizable improvement of the leptoquark mass and coupling limits is
expected from high luminosity LHC data.
Exclusion limits expected from the leptoquark pair production
at LHC will extend up to the leptoquark masses of about 2.2 TeV.
The possible $S_{1}$ and $\tilde{V}_{\circ}$ leptoquark ``signal'', 
resulting from the new data on the atomic parity violation, 
can be confirmed or excluded even for very high leptoquark masses.
If the leptoquark type particle is discovered at LHC, with mass below 1 TeV,
TESLA and THERA will be an ideal place to study its properties, provided
that the Yukawa coupling is not too small.
For $e\gamma$ scattering at TESLA leptoquarks with Yukawa couplings
down to  $\lambda_{LQ}\sim$ 0.01 can be studied.

%
\section*{Acknowledgements}

This work has been partially supported by the Polish State Committee 
for Scientific Research (grant No. 2 P03B 035 17).

%
%



\begin{thebibliography}{}

\bibitem{myglqa}   
A.F.\.Zarnecki hep-ph/0003271.

\bibitem{brw}   
W.Buchm\"uller, R.R\"uckl and D.Wyler, Phys. Lett. \textbf{B191} (1987) 442; \\
Erratum: Phys. Lett. \textbf{B448} (1999) 320.

\bibitem{aachen} 
A.Djouadi, T.K{\"o}hler, M.Spira, J.Tutas,  Z. Phys. \textbf{C46}{679}{1990}.

\bibitem{lqci}  
J.Kalinowski, R.R\"uckl, H.Spiesberger and P.M.Zerwas, 
                        Z. Phys. \textbf{C74} (1997) 595.

\bibitem{apvnew}    
S.C.Bennett and C.E.Wieman, Phys. Rev. Lett. \textbf{82} (1999) 2484.

\bibitem{lepnew}    
LEP Electroweak Working Group, C.Geweniger {\it et al}, LEP2FF/00-01.

\bibitem{compton} 
I.F.Ginzburg, G.L.Kotkin, V.G.Serbo and V.I.Telnov,  \\
Nucl. Inst. Meth. \textbf{205} (1983) 47.

\bibitem{mygcia}   
A.F.\.Zarnecki Euro. Phys. J. \textbf{C11} (1999) 539; \\
A.F.\.Zarnecki hep-ph/0006196.

\bibitem{cdfdy}    
The CDF Collaboration, F.Abe et al., 
             Phys. Rev. Lett \textbf{79} (1997) 2198;\\
The CDF Collaboration, F.Abe et al., 
             Phys. Rev. \textbf{D59} (1999) 052002.

\bibitem{cmsdy}    
A.K.Gupta, S.Jain, N.K.Mondal, CMS Note 1999/075.

\bibitem{mrst} 
A.D. Martin, R.G. Roberts, W.J. Stirling, R.S Thorne, \\
Euro. Phys. J. \textbf{C4} (1998) 463.

\bibitem{grv}  
M.Gl\"uck, E.Reya and A.Vogt, Phys. Rev. \textbf{D46} (1992) 1973.

\bibitem{jb1}  
J.Bl\"umlein, E.Boos, A.Kryukow,  Z. Phys. \textbf{C76} (1997) 137.

\bibitem{jb2}   
J.Bl\"umlein, E.Boos, A.Kryukow, DESY 97-067.

\bibitem{cmslq}    
S.Abdullin, F.Charles, F.Luckel, CMS Note 1999/027.

\bibitem{darin}  
D.Acosta, S.Blessing, Annu. Rev. Nucl. Part. Sci. \textbf{49} (1999) 389.

\bibitem{egdir}  
F.Cuypers, Nucl. Phys. \textbf{B474} (1996) 57.

\end{thebibliography}
\end{document}